\documentclass[graphicx, 12pt]{article}
\usepackage{indentfirst}

\usepackage{graphicx}
\usepackage{caption}
\usepackage{subfigure}
\usepackage{ctex,amsmath,amssymb}
\usepackage[square, comma, sort&compress, numbers]{natbib}

\begin{document}

\small
\hoffset=-1truecm
\voffset=-2truecm
\title{\bf The analytical discussion on strong gravitaional lensing
for a gravitational source with a $f(R)$ global monopole}
\author{Jingyun Man \hspace {1cm} Hongbo Cheng\footnote {E-mail address: hbcheng@ecust.edu.cn}\\
Department of Physics, East China University of Science and
Technology,\\ Shanghai 200237, China}

\date{}
\maketitle

\begin{abstract}
Here the gravitational lensing in strong field limit of a
Schwarzschild black hole with a solid deficit angle owing to
global monopole within the context of the $f(R)$ gravity theory is
investigated. We obtain the expressions of deflection angle and
time delay in the forms of elliptic integrals and discuss the
asymptotic behaviour of the elliptic integrals to find the
explicit formulae of angle and time difference in the strong field
limit. We show that the deflection angle and the time delay
between multiple images are related not only to the monopole but
also to the $f(R)$ correction $\psi_{0}$ by taking the
cosmological boundary into account. Some observables such as the
minimum impact parameter, the angular separation, the relative
magnification and the compacted angular position have been
estimated as well. It is intriguing that the tiny modification on
the standard general relativity will make the remarkable deviation
on the angle and the time lag, offering a significant way to
explore some possible distinct signatures of the topological
soliton and the correction of the Einstein general relativity.
\end{abstract}
\vspace{4cm} \hspace{1cm} PACS number(s): 04.70.Bw, 14.80.Hv

\section{ Introduction }

Gravitational lensing is an important astrophysical application of
general relativity and a powerful probe to gravitational source,
lens object and spacetime structure [1-6]. We can make use of the
gravitational lensing to investigate the distant stars no matter
they are bright or dim. If the lens is massive enough like a black
hole, electromagnetic radiation can approach very close to the
object while the deflection angle will exceed $3\pi/2$ [7] and it
encodes the information from strong field caused by a compact
body. In this circumstances, a sequence of images are formed on
both side of the optic axis, which are called as relativistic
images, due to large deflections of light more than $2\pi $ apart
from the so-called primary and secondary images observed in weak
gravitational field and formed due to small deflection of light
rays \cite{Vivbhadra62}.

In general, the deflection angle of photons passing close to a
compact and massive source is expressed in integral forms, so it
is difficult to discuss the detailed relation between the angle
and the gravitational source or the spacetime geometry.
Alternatively we perform the calculation of the integral
expression in strong field limit where the minimum distance a
photon is able to approach to the black hole. The analytic method
proposed by Bozza is to expand the integral expression towards the
photo sphere \cite{Bozza66,Capozziello33,Mutka581,Mahonen576,Beloborodov566,Keeton72} which showed that there exists a logarithmic
divergence of the deflection angle in the proximity of the photon
sphere. The strong gravitational lensing was treated in a
Schwarzschild black hole and a Schwarzschild black hole and a
Reissner-Nordstr\"{o}m black hole \cite{Bozza66}, a GMGHS charged black hole
\cite{Bhadra67}, a spinning black hole \cite{Bozza67}, a braneworld black hole \cite{Whisker71},
the deformed Horava-Lifshitz black hole \cite{Chen80}, the black hole with
a global monopole \cite{Cheng28}. In addition to Bozza's scheme, the
explicit calculation of elliptic integrals is also valid and
powerful and was used in the description of strong deflection of
massive particle around the supermassive black hole \cite{Tsupko89}.

If photons propagates from the emitter to the observers along
different rays, then the light-travel time corresponding to every
image is obvious different. The lag of time between the multiple
images is called time delay. Generally, the travel time is not
observable. However, in the situation of the appearance of
multiple images, the time delay can be observed if the intrinsic
luminosity of the source is time-dependent. Therefore, the
luminosity variations can be used to describe the geometry of the
lens which is related with the images as a relative temporal
phase. As an advantage, time delay is a one-dimensional quantity
and can be used to test the underlying cosmological expansion \cite{Schneider1992}.
The measurement of time delay provides a probe of the Hubble
constant \cite{Refsdal128,Oguri660}. The general approximative expressions of time
delay between relativistic images in strong field limits for
asymptotic flat spacetime without a cosmological horizon has been
presented in \cite{Mancini36}.

Several types of topological objects such as domain walls, cosmic
strings and monopole may have been formed during the vacuum phase
transition in the early Universe \cite{Kibble9,Vilenkin121}. These topological
defects appeared due to breakdown of local or global gauge
symmetries. A global monopole is a spherical symmetric topological
defect formed in the phase transition of a system composed by a
self-coupling triplet of scalar field whose original global O(3)
symmetry is spontaneously broken to U(1). The properties of the
metric outside a monopole are investigated in \cite{Barriola63}, which also
show that the monopole exerts practically no gravitational force
on nonrelativistic matter, but the space around it has a solid
angle and all light rays are deflected at the same angle. We have
considered the gravitational lens equation for the massive global
monopole in the strong field limit to exhibit the correlation
between the deflection angle and the deficit solid angle subject
to the monopole model parameters in the standard general
relativity \cite{Cheng28}.

However, subject to the fact of the accelerated expansion of the
universe, the metric with $f(R)$ modification describes the
spacetime more completely. The theory of $f(R)$ gravity is a type
of modified gravity theory first proposed by Buchdahl \cite{Buchdahl150}, and
has been applied to explain the accelerated-inflation problem
instead of adding dark energy or dark matter \cite{Nojiri68,Carrol70,Fay74}.  The
gravitational field of a global monopole in the modified gravity
theory has been discussed \cite{Carames3}. Further, in \cite{Carames27}, they find that
the presence of the parameter associated with the modification of
gravity is the indispensable to provide stable circular orbits for
particles. The nonvanishing modified parameter $\psi_{0}$ also
bring a cosmological horizon as a boundary of universe to the
spacetime described by the $f(R)$ monopole metric, but the
spacetime without gravity modification is asymptotic flat. It should be notice that the asymptotic flat spacetime is an essential condition for derivation of gravitational lensing in both weak field limits \cite{Weinberg} and strong field limits \cite{Bozza66}. Here we use the elliptic integrals to rewrite the expressions of the deflection angle and the time delay which contain polynomials for three or higher order. This method further present the analytic results when the test particle close to the photon sphere whether the size and scale of the observable universe exist or not.

In this paper, we plan to probe the strong gravitational lensing
in terms of deflection angle of lights and time delay of multiple
images in strong field limit on the massive source swallowing a
global monopole governed by $f(R)$ theory. In next section we give
a brief introduction about the metric considered here. In section
III, the integral form of deflection angle of light ray is
derived. We discuss the asymptotic behaviour of the elliptic
integrals to present the expression of deflection angle at the
position close to the photon sphere, and we perform the numerical
estimation of observables as well. In section IV, we put forward
the time delay in this background with double horizons as elliptic
integrals. Further we calculate the strong-limit time delay by
means of the series representations of elliptic integrals.
Finally, we discuss our result in section V.

\section{The Schwarzschild black hole with a $f(R)$ global monopole}

The simplest model that gives rise to global monopole is describe
by the Lagrangain [26],

\begin{equation}
{\mathcal{L}} =\frac{1}{2} (\partial_{\mu}\phi
^{a})(\partial^{\mu}\phi ^{a})-\frac{1}{4} \lambda
(\phi^{a}\phi^{a}-\eta^{2})^{2}.
\end{equation}

\noindent The triplet of field configuration showing a monopole is

\begin{equation}
\phi ^{a}=\eta h(r)\frac{x^{a}}{r},
\end{equation}

\noindent where $x^{a}a^{a}=r^{2}$. Here $\lambda$ and $\eta $ are
model parameter. This model has a global O(3) symmetry, which is
spontaneously broken to U(1). In order to couple this matter field
to the gravitational field equation in the $f(R)$ theory and
obtain their spherically symmetric solution, we adopt the line
element as follow,

\begin{equation}
ds^{2}=A(r)dt^{2}-B(r)dr^{2}-r^{2}(d\theta ^{2}+sin^{2}\theta d\varphi^{2}).
\end{equation}

In the $f(R)$ gravity theory, the action is given by \cite{Carames3},

\begin{equation}
S=\frac{1}{2\kappa}\int d^{4}x\sqrt{-g}f(R)+S_{m},
\end{equation}

\noindent where $f(R)$ is an analytical function of Ricci scalar
$R$ and $\kappa=8\pi G$. $G$ is the Newton constant. $g$ is the
determinant of metric tensor. $S_{m}$ is the action associated
with the matter fields. According to the metric formalism, the
field equation leads,

\begin{equation}
F(R)R_{\mu\nu}-\frac{1}{2}f(R)g_{\mu\nu}-\nabla_{\mu}\nabla_{\nu}F(R)+g_{\mu\nu}  F(R)=\kappa T_{m\mu\nu},
\end{equation}

\noindent where $F(R)=df(R)/dR$ and $T_{m\mu\nu}$ is the minimally
coupled energy-momentum tensor. Under the weak field approximation
that assumes the components of metric tensor like $A(r)=1+a(r)$
and $B(r)=1+b(r)$ with $|a(r)|$ and $|b(r)|$ being smaller than
unity, the field equation is solved in \cite{Carames27}. The metric is found
finally,

\begin{equation}
A(r)=B^{-1}(r)=1-8\pi G\eta^{2}-\frac{2GM}{r}-\psi_{0}r.
\end{equation}

\noindent Here the modification theory of gravity corresponds to a
small correction on the general relativity like
$F(R(r))=1+\psi(r)$ with $\psi(r)\ll 1$. It can also be taken as
the simplest analytical function of the radial coordinate
$\psi(r)=\psi_{0}r$. In this case the factor $\psi_{0}$ reflects
the deviation of standard general relativity. Here the correction
$\psi_{0}r$ in the metric is linear, which is different from those
in the case such as de Sitter spacetime and Reissner-Nordstr\"{o}m
metric etc.. It should be pointed that for a typical grand unified
theory the monopole parameter $\eta$ is of the order $10^{16}GeV$,
which means $8\pi G\eta^{2}\approx 10^{-5}$. The mass parameter is
$M\sim M_{core}$ which is very small.

We choose that both the observer and the gravitational source lie
in the equatorial plane with condition $\theta =\frac{\pi}{2}$.
The whole trajectory of the photon is limited to the same plane.
On the equatorial plane the metric reads,

\begin{equation}
ds^{2}=A(r)dt^{2}-B(r)dr^{2}-C(r)d\varphi^{2},
\end{equation}

\noindent where

\begin{equation}
C(r)=r^{2}.
\end{equation}

\noindent
We note that with the presence of a nonzero $\psi_{0}$ a cosmological horizon,

\begin{equation}
r_{c}=\frac{1}{\psi_{0}}\left(1-8\pi G\eta^{2}+ \sqrt{(1-8\pi G\eta^{2})^{2}-8GM\psi _{0}}\right),
\end{equation}

\noindent appears, besides an event horizon,

\begin{equation}
r_{h}=\frac{1}{\psi_{0}}\left(1-8\pi G\eta^{2}- \sqrt{(1-8\pi G\eta^{2})^{2}-8GM\psi _{0}}\right).
\end{equation}

\noindent The nonvanishing modified parameter $\psi_{0}$ also
bring a cosmological horizon as a boundary of universe to the
spacetime described by the $f(R)$ monopole metric, but the
spacetime without gravity modification is asymptotic flat.

\section{The deflection angle of a massive source with a
$f(R)$ global monopole}

The deflection angle for the electromagnetic ray coming from the
source to the observer can be expressed as a function of the
closest approach \cite{Virbhadra337},

\begin{equation}\label{angle}
\alpha=I(r_{0})-\pi
\end{equation}

\noindent where

\begin{equation}
I(r_{0})=I_{OL}(r_{0})+I_{LS}(r_{0})=\int_{r_{0}}^{D_{OL}}\left|
\frac{d\varphi}{dr}\right| dr+\int_{r_{0}}^{D_{LS}}\left|\frac{d\varphi}{dr}\right|dr,
\end{equation}

\noindent and

\begin{equation}
\frac{d\varphi}{dr}=\frac{\sqrt{B(r)}}{\sqrt{C(r)}\sqrt{\frac{C(r)}{C(r_{0})}\frac{A(r_{0})}{A(r)}-1}}.
\end{equation}

\noindent Here $r_{0}$ is the minimum distance from the photon
path to the source, $D_{OL}$ is the distance of the lens from the
observer and $D_{LS}$ is the distance of the lens from the source.
We should note that $r_{0}<D_{OL}<r_{c}$ and $r_{0}<D_{LS}<r_{c}$.
It requires that the deflection angle turn to be infinite, meaning
that the denominator of expression (13) is equal to the zero. To
achieve this aim, we solve the equation
$\frac{C'(r)}{C(r)}=\frac{A'(r)}{A(r)}$ to obtain the radius of
the photon sphere \cite{Virbhadra65, Claudel42}. Certainly the
closest approach distant r0 must be larger than the radius of
photon sphere or the photon will move around the gravitational
source forever instead if escaping from the source. The radius of
the photon sphere in the $f(R)$ global monopole metric is given
by,

\begin{eqnarray}
r_{m}&&=\frac{1-8\pi G\eta^{2}-\sqrt{(1-8\pi G\eta^{2})^{2}-6GM\psi_{0}}}{\psi_{0}}\nonumber\\
&&\approx \frac{3GM}{1-8\pi G\eta ^{2}}+O(\psi_{0}).
\end{eqnarray}

\noindent When we neglect the influence from $f(R)$ theory
$\psi_{0}=0$, the photon sphere radius (14) will recover to that
of metric by massive object involving global monopole within the
frame of Einstein's general relativity \cite{Cheng28}. It can be
checked as $r_{h}<r_{m}<r_{c}$, which indicates that a photon
sphere will survive for the spacetime with two horizons.

For the conservation of angular momentum, at $r=r_{0}$, we define
the impact parameter related with the minimum approach by
\cite{Virbhadra337},

\begin{equation}
y=\sqrt{\frac{C(r_{0})}{A(r_{0})}}
\end{equation}

\noindent The angular separation can be approximately expressed by
$\theta=\frac{y}{D_{OL}}$. The minimum impact parameter
corresponding to the radius of the photon sphere will thus be

\begin{eqnarray}
y_{m}=\sqrt{\frac{r_{m}^{3}}{-\psi_{0}r_{m}^{2}+(1-8\pi G\eta^{2})
r_{m}-2GM}}.
\end{eqnarray}

Now we rewrite the integral expression for the deflection angle
(16) with the help of elliptic functions. At first we introduce
the notation like $u=\frac{1}{r}$ leading,

\begin{equation}
 u_{0}=\frac{1}{r_{0}}, u_{m}=\frac{1}{r_{m}}, u_{OL}=\frac{1}{r_{OL}}, u_{LS}=\frac{1}{r_{LS}}.
\end{equation}

\noindent The deflection angle (\ref{angle}) becomes,

\begin{eqnarray}\label{alpha}
\alpha=&&\int_{u_{OL}}^{u_{0}}\frac{du}{\sqrt{2GM(u-u_{0})
(u-u_{1})(u-u_{2})}}\nonumber\\
&&+\int_{u_{LS}}^{u_{0}}\frac{du}{\sqrt{2GM(u-u_{0})
(u-u_{1})(u-u_{2})}}-\pi,
\end{eqnarray}

\noindent where

\begin{eqnarray}
u_{1,2}=&&\frac{1}{4GM}\Bigg(1-8\pi G\eta^{2}-2GMu_{0}\hspace{7cm}\nonumber\\
&&\pm\sqrt{(1-8\pi G\eta^{2})^{2}+4(1-8\pi
G\eta^{2})GMu_{0}-8GM\psi_{0}-12(GMu_{0})^{2}}\Bigg).
\end{eqnarray}

\noindent where the upper sign "+" belongs to $u_{1}$ and the
lower ones "-" is for $u_{2}$. According to Ref. \cite{Gradshteyn}, the two
integral parts of Eq. (\ref{alpha}) can be written as,

\begin{equation}
\int_{u_{LS}(u_{OL})}^{u_{0}}\frac{du}{\sqrt{2GM(u-u_{0})
(u-u_{1})(u-u_{2})}}=\frac{1}{\sqrt{2GM}}\frac{2}{\sqrt{u_{1}-u_{2}}}
F(\delta_{LS(OL)}, q),
\end{equation}

\noindent Here $F(\delta_{LS(OL)}, q)$ is an elliptic integral of the first
kind \cite{Gradshteyn},

\begin{eqnarray}
F(\delta_{LS(OL)},
q)=\int_{0}^{\delta_{LS(OL)}}\frac{d\alpha}{\sqrt{1-q^{2}\sin^{2}\alpha}}\nonumber\\
=\int_{0}^{\sin\delta_{LS(OL)}}\frac{dx}{\sqrt{(1-x^{2})(1-q^{2}x^{2})}},
\end{eqnarray}

\noindent where

\begin{equation}\label{sin}
\sin\delta_{LS(OL)}=\sqrt{\frac{(u_{1}-u_{2})(u_{0}-u_{LS(OL)})}
{(u_{0}-u_{2})(u_{1}-u_{LS(OL)})}},
\end{equation}

\begin{equation}\label{q}
q=\sqrt{\frac{u_{0}-u_{2}}{u_{1}-u_{2}}}.
\end{equation}

When the photon-travel paths near the source, the deflection angle
will be bigger. If the angle is greater than $2\pi$, the photon
will circle the massive source several times before they reach the
observers. When the minimum distance from the photon-travel paths
to the source $r_{0}$ approaches to the radius of photon sphere,
the parameters (\ref{sin}) and (\ref{q}) will be,

\begin{equation}
\sin\delta_{LS(OL)}|_{u_{0}=u_{m}}=1,
\end{equation}

\noindent and

\begin{equation}
q|_{u_{0}=u_{m}}=1.
\end{equation}

\noindent The asymptotic behaviour of an elliptic integral of the
first kind is \cite{Gradshteyn},

\begin{equation}
\lim_{q\longrightarrow1}F(\delta_{LS(OL)},q)=\ln\frac{4}{\sqrt{1-q^{2}}}
-\ln\cot\frac{\delta_{LS(OL)}}{2}+O(1-q^{2}).
\end{equation}

\noindent The result is independent on the position of the source or the observer under the strong field condition, thus $I_{OL}(r_{0}\rightarrow r_{m})=I_{LS}(r_{0}\rightarrow r_{m})$.

 Within the region just containing the photon sphere, we
expand the deflection angle expression (19) in virtue of the
properties of elliptic functions,

\begin{equation}\label{deflection angle}
\alpha(\theta)=-aln(\frac{\theta D_{OL}}{y_{m}}-1)+b+O(y-y_{m}),
\end{equation}

\noindent where the coefficients of the deflection angle are

\begin{equation}
a=\frac{1}{[(1-8\pi G\eta^{2})^{2}-6GM\psi_{0}]^{\frac{1}{4}}},
\end{equation}

\begin{eqnarray}
b=2a\Bigg[\frac{1}{2}ln\sigma+3ln2+ln3-ln\frac{1-8\pi G\eta^{2}+\sqrt{(1-8\pi G\eta^{2})^{2}-6GM\psi_{0}}}{\sqrt{(1-8\pi G\eta^{2})^{2}-6GM\psi_{0}}}\Bigg]-\pi.
\end{eqnarray}

\noindent Here we have used,

\begin{equation}
\frac{r_{0}}{r_{m}}-1=\left[\frac{1}{\sigma}(\frac{y}{y_{m}}-1)\right]^{\frac{1}{2}},
\end{equation}

\noindent where

\begin{equation}
\sigma=\frac{12G^{2}M^{2}-20GM\psi_{0}r_{m}^{2}+4(1-8\pi G\eta^{2})r_{m}^{3}\psi_{0}-\psi_{0}^{2}r_{m}^{4}}{8[-(1-8\pi G\eta^{2}) r_{m}+2GM+\psi_{0}r_{m}^{2}]^{2}}.
\end{equation}

\noindent
Figure. 1 shows the deflection angle
in the strong field limit, $y=y_{m}+0.003GM$, for various values of $\psi_{0}$ and $\eta$. We see that, $\alpha$ increases as $GM\psi_{0}$ increases.

We relate the position and the magnification to the strong field limit coefficients for the sake
of comparing our results with the observable evidence. The position of the source and the images
are related through the lens equation derived in \cite{Capozziello33} given by

\begin{equation}
\beta=\theta-\frac{D_{LS}}{D_{OL}}\bigtriangleup \alpha_{n},
\end{equation}

\noindent where $\beta$ denotes the
angular separation between the source and the lens, and $\theta$
is the angular separation between the lens and the image. The
offset of the deflection angle is expressed as $\bigtriangleup
\alpha_{n}=\alpha(\theta)-2n\pi$ by subtracting all the times run
around the source by photons. Due to $y_{m}\ll D_{OL}$ the
position of the n-th relativistic image can be approximated as,

\begin{equation}\label{theta}
\theta_{n}=\theta_{n}^{0}+\frac{y_{m}e_{n}(\beta-\theta_{n}^{0})D_{OS}}{a D_{LS}D_{OL}},
\end{equation}

\noindent where

\begin{equation}
e_{n}=exp(\frac{b-2n\pi}{a}),
\end{equation}

\noindent and $D_{OS}=D_{OL}+D_{LS}$, while the second term in the right-hand side of Eq. (\ref{theta})
is much smaller than $\theta_{n}^{0}$ and we introduce
$\theta_{n}^{0}$ as $\alpha(\theta_{n}^{0})=2n\pi$. The
magnification of n-th relativistic image is the inverse of the
Jacobian evaluated at the position of the image and is obtained
as,

\begin{equation}
y_{n}=\frac{y_{m}e_{n}(1+e_{n})D_{OS}}{a\beta D_{LS}D_{OL}^{2}}.
\end{equation}

\noindent In the limit $n\rightarrow \infty$ the asymptotic
position of approached by a set of images $\theta_{\infty}$
relates to the minimum impact parameter as,

\begin{equation}
y_{m}=D_{OL}\theta_{\infty}.
\end{equation}

As an observable the angular separation between the first image
and the others is defined as,

\begin{equation}
s=\theta_{1}-\theta_{\infty}=\theta_{\infty}e^{\frac{b-2\pi}{a}},
\end{equation}

\noindent where $\theta_{1}$ represents the outermost image in the
situation that the outermost one is thought as a single image and
all the remaining ones are packed together at $\theta_{\infty}$.
The ratio of the flux from the first image and the flux of all the
other images is,

\begin{equation}
{\mathcal{R}} =\frac{\mu_{1}}{\sum\limits_{n=2}^{\infty}\mu_{n}}=e^{\frac{2\pi}{a}}.
\end{equation}

\noindent According to $e^{\frac{2\pi}{a}}\ll 1$ and
$e^{\frac{b}{a}}\sim 1$, these observables can be written in terms
of the deflection angle parameters, which is presented in Eqs.
(42) and (43). As another observable the magnification of the
first image with the other images can be thus defined as
${\mathcal{R}}_{m}=2.5$log${\mathcal{R}}$. Take the black-hole
mass $M=2.8\times 10^{6}M_{\odot}$ in the center of our galaxy and
$D_{OL}=8.5kpc$ as the distance between the sun and the galaxy
center.

It is important that the strong field limit coefficients such as
$a$, $b$ and $y_{m}$ are directly connected to the observables
like ${\mathcal{R}}_ {m}$ and $s$. It is then possible for us to
probe whether the original general relativity need to be
generalized in virtue of the strong field gravitational lensing
for a Schwarzschild black hole with a global monopole. From Figure
2 and Figure 3, with increasing the monopole parameter, all curves
of the angular separation $s$ and of the asymptotic position of
the set of outer images $\theta_{\infty}$ rise while the
${\mathcal{R}}_ {m}$ curves decrease. In Figure 4, the
magnification of relativistic images decreases when $GM\psi_{0}$
increases and when $8\pi G\eta^{2}$ increases, which means the
difference of the flux from the first image and the flux of all
the other images reduces due to the modification of gravity
theory. We present the estimations in Table 1 to show how the
appearance of the deviation of standard general relativity
enhances the observables for strong gravitational lensing. For
example, the angular separation between the first image and all
the other images for $f(R)$ Schwarzschild lens is $310$ times
larger than the value for Schwarzschild lens, which means it is
unambiguous to distinguish the relativistic images. Once
observational devices catch such multiple images, it is the
indispensable evidence to study the $f(R)$ gravity. Hence, the
strong gravitational lensing for massive source with a global
monopole is an efficient probe to enlarge the effect of the
deviation owing to the correction to the Einstein's general
relativity although the correction itself is small.

\begin{table}[h!]
    \begin{center}
        \newcommand{\tabincell}[2]{\begin{tabular}{@{}#1@{}}#2\end{tabular}}
        \begin{tabular}{c|c|ccccccc}\hline
         & Schwarzschild &\multicolumn{6}{c}{$f(R)$ ($\eta=0$)}\\\hline
        $GM\psi_{0}$ & 0 & 0.001 & 0.01 & 0.05 & 0.08 & 0.1 & 0.11 & 0.12\\
        $\theta_{\infty}$ ($\mu$ $arc$ $sec$) & 16.8 & 16.9 & 17.6 & 22.3 & 29.4 & 40.1 & 52.1 & 91.1 \\
        s ($\mu$ $arc$ $sec$) & 0.29 & 0.30 & 0.36 & 0.90 & 2.43 & 6.90 & 16.06 & 90.15 \\
        $\frac{y_{m}}{GM}$ & 5.196 & 5.220 & 5.445 & 6.896 & 9.080 & 12.369 & 16.103 & 28.147 \\
        $\mathcal{{R}}_ {m}$ & 6.82 & 6.46 & 6.35 & 5.76 & 5.17 & 4.62 & 4.25 & 3.74\\\hline

        \end{tabular}
        \caption{\label{tab:{table1}} Estimations for the variable observables such as
         the compacted angular position, the angular separation and magnification
         of the first image and the other images, and the minimum impact parameter
         for a Schwarzschild lens in the center of Milk Way or for the lens
         with $f(R)$ modification. The global monopole parameter is zero.}
    \end{center}
\end{table}

\section{The time delay in strong gravitational lensing for the
massive source with a $f(R)$ global monopole}

For equatorial geodesics in sphere symmetric spacetime, the
equation for time and radial position for the motion of photons
around the gravitational source is given by \cite{Virbhadra77}

\begin{equation}
\frac{dt}{dr}=\pm\frac{\sqrt{B(r)}}{\sqrt{A(r)}\sqrt{1-\frac{A(r)C(r_{0})}{A(r_{0})C(r)}}}.
\end{equation}

\noindent The duration that a photon emitted by a star or cluster
passes the lens and reaches the receiver is,

\begin{eqnarray}\label{time}
T=\int_{D_{LS}}^{D_{OL}}\frac{dt}{dr}dr\hspace{2cm}\nonumber\\
=\int_{r_{0}}^{D_{LS}}\Bigg|\frac{dt}{dr}\Bigg|dr+\int_{r_{0}}^{D_{OL}}
\Bigg|\frac{dt}{dr}\Bigg|dr.
\end{eqnarray}

\noindent We substitute the metric (8) into Eq. (\ref{time}) to find that,

\begin{eqnarray}\label{time part}
\int_{r_{0}}^{D_{LS}(D_{OL})}\Bigg|\frac{dt}{dr}\Bigg|dr\hspace{8cm}\nonumber\\
=\frac{\sqrt{(1-8\pi
G\eta^{2})u_{0}^{2}-2GMu_{0}^{3}-\psi_{0}u_{0}}}{(2GM)^{\frac{3}{2}}}
\hspace{5cm}\nonumber\\
\times\int_{u_{0}}^{u_{LS}(u_{OL})}\frac{du}{u(u-u_{3})(u-u_{4})
\sqrt{(u-u_{0})(u-u_{1})(u-u_{2})}},
\end{eqnarray}

\noindent with

\begin{equation}
u_{LS(OL)}=\frac{1}{D_{LS(OL)}},
\end{equation}

\noindent while

\begin{equation}
u_{3,4}=\frac{1}{4GM}[(1-8\pi G\eta^{2})\pm\sqrt{(1-8\pi
G\eta^{2})^{2}-8GM\psi_{0}}].
\end{equation}

\noindent where the sign "+" and "-" subject to $u_{3}$ and
$u_{4}$ respectively. According to Ref. \cite{Gradshteyn}, the
integral parts of Eq. (\ref{time part}) can be rewritten as,

\begin{eqnarray}
\int_{u_{0}}^{u_{LS}(u_{OL})}\frac{du}{u(u-u_{3})(u-u_{4})
\sqrt{(u-u_{0})(u-u_{1})(u-u_{2})}}\hspace{2.5cm}\nonumber\\
=\frac{2}{u_{3}(u_{3}-u_{4})(u_{1}-u_{3})(u_{0}-u_{3})
\sqrt{u_{1}-u_{2}}}\hspace{5cm}\nonumber\\
\times[(u_{0}-u_{1})\Pi(\delta_{LS(OL)},
q^{2}\frac{u_{3}-u_{1}}{u_{3}-u_{0}},q)+(u_{3}-u_{0})
F(\delta_{LS(OL)},q)]\nonumber\\
+\frac{2}{u_{3}u_{4}u_{1}u_{0}\sqrt{u_{1}-u_{2}}}\hspace{8cm}\nonumber\\
\times[(u_{0}-u_{1})\Pi(\delta_{LS(OL)},
q^{2}\frac{u_{1}}{u_{0}},q)+(-u_{0})
F(\delta_{LS(OL)},q)]\hspace{1cm}\nonumber\\
-\frac{2}{u_{4}(u_{3}-u_{4})(u_{1}-u_{4})(u_{0}-u_{4})
\sqrt{u_{1}-u_{2}}}\hspace{4.5cm}\nonumber\\
\times[(u_{0}-u_{1})\Pi(\delta_{LS(OL)},
q^{2}\frac{u_{4}-u_{1}}{u_{4}-u_{0}},q)+(u_{4}-u_{0})
F(\delta_{LS(OL)},q)].
\end{eqnarray}

\noindent Here $\Pi(\delta, n, q)$ is an elliptic integral of the
third kind \cite{Gradshteyn},

\begin{eqnarray}
\Pi(\delta, n, q)=\int_{0}^{\delta}\frac{d\alpha}
{(1-n\sin^{2}\alpha)\sqrt{1-q^{2}\sin^{2}\alpha}}\nonumber\\
=\int_{0}^{\sin\delta}\frac{dx}{(1-nx^{2})\sqrt{(1-x^{2})(1-q^{2}x^{2})}}.
\hspace{1cm}
\end{eqnarray}

\noindent In the case of large deflection angle,

\begin{equation}
\sin\delta_{LS(OL)}|_{u_{0}=u_{m}}=1.
\end{equation}

\noindent while,

\begin{equation}
\lim_{r_{0}\longrightarrow r_{m}}\int_{r_{0}}^{D_{LS}}
|\frac{dt}{dr}|dr=\lim_{r_{0}\longrightarrow
r_{m}}\int_{r_{0}}^{D_{OL}} |\frac{dt}{dr}|dr
\end{equation}

\noindent Hence,

\begin{equation}
T=-\bar{a}ln(\frac{y}{y_{m}}-1)+\bar{b}+O(y-y_{m}),
\end{equation}

\noindent where

\begin{eqnarray}
\bar{a}=&&\frac{\sqrt{u_{m}[(1-8\pi G\eta^{2}) u_{m}-2GMu_{m}^{2}
-\psi_{0}]}}{(2GM)^{\frac{3}{2}}\sqrt{u_{m}-u_{2m}}}\\\nonumber
&& \times\Bigg(\frac{-1}{u_{m}u_{3}u_{4}}+\frac{1}{u_{3}(u_{3}-u_{4})
(u_{3}-u_{m})}-\frac{1}{u_{4}(u_{3}-u_{4})(u_{4}-u_{m})}\Bigg),
\end{eqnarray}

\noindent and $\bar{b}$ is a contant irrelevant to the impact
parameter belong to the variance relativistic images. In strong
gravitational field limit for a Schwarzschild black hole with a
$f(R)$ global monopole,  the time delay of two images on the same
side of the lens is given,

\begin{equation}\label{time delay}
\Delta T_{n,m}^{s}=2\pi y_{m}(n-m),
\end{equation}

\noindent and for the two images lying on the opposite side of lens,

\begin{equation}
\bigtriangleup T_{n,m}^{o}=2y_{m} \left[\pi(n-m)-\gamma\right].
\end{equation}

\noindent where

\begin{eqnarray}
\frac{\bar{a}}{a}=y_{m},
\end{eqnarray}

\noindent and $n$ and $m$ are different times of photons winding
around the black hole, $\gamma$ is the angular separation between
the source and the optical axis.. The expression of deflection
angle (\ref{deflection angle}) has been considered to obtian Eq.
(\ref{time delay}). More commonly, if the source are highly
aligned with the lens, the gravitational lensing effects become
more prominent so that $\gamma\sim D^{-1}_{OL}<<2\pi$ [22, 34].
Then Eqs. (50), (51) are reduced to the same result provided in
Ref. \cite{Mancini36} if we recover the physical units and
consider the time delay between first image,

\begin{equation}
\bigtriangleup T_{2,1}=\frac{2\pi}{c}D_{OL}\theta_{\infty},
\end{equation}

\noindent where $c$ is the speed of light. However, we deduce Eqs.
(50),(51) under strong field approximation using elliptic integral
without rejection of exponential terms \cite{Mancini36}. It should
be notice that the analytical results from Eq. (53) were shown to
have large percentage errors in \cite{Virbhadra77}. We present our
results in Table 2. Since

\begin{equation}
\frac{\bigtriangleup T^{f(R)}_{2,1}}{\bigtriangleup T^{Sch}_{2,1}}=\frac{y^{f(R)}_{m}}{y^{Sch}_{m}}
\end{equation}

\noindent where the superscripts $f(R)$ and $Sch$ represent the
case of spacetime with $f(R)$ modification involved and
Schwarzschild spacetime, we find $\bigtriangleup
T^{f(R)}_{2,1}=5.4\times\bigtriangleup T^{Sch}_{2,1}$, if
$GM\psi_{0}=0.12$; When $GM\psi_{0}=0.11$, $\bigtriangleup
T^{f(R)}_{2,1}=3.1\times\bigtriangleup T^{Sch}_{2,1}$; When
$GM\psi_{0}=0.01$, $\bigtriangleup
T^{f(R)}_{2,1}=1.05\times\bigtriangleup T^{Sch}_{2,1}$. From
Figure 5, either the deviation from the standard general
relativity or the topological defect can enhance the time delay
although the deviations are fairly tiny.

\begin{table}[h!]
\begin{center}
\newcommand{\tabincell}[2]{\begin{tabular}{@{}#1@{}}#2\end{tabular}}
\begin{tabular}{c|c|c|c|cc}\hline
\tabincell{c}{black hole \\ in galaxy} & \tabincell{c} {Mass \\ ($M_{\bigodot}$)} & \tabincell{c}{Distance \\ (Mpc)} & \tabincell{c} {Schwarzschild \\ $\bigtriangleup T^{Sch}_{2,1}$(min)} & \tabincell{c} {$GM\psi_{0}=0.12$ \\ $\bigtriangleup T^{0.12}_{2,1}$(min)} & \tabincell{c} {$GM\psi_{0}=0.1$ \\ $\bigtriangleup T^{0.1}_{2,1}$(min)}  \\ \hline
Milk Way & $2.8\times10^{6}$ & 0.0085 & 7.5 & 40.6 & 17.9 \\
NGC4486(M87) & $3.3\times 10^{9}$ & 15.3 & 8839.3 & 47880.5 & 21040.4 \\
NGC3115 & $2.0\times 10^{9}$ & 8.4 & 5357.1 & 29018.5 & 12751.8
\\
NGC4374(M84) & $1.4\times 10^{9}$ & 15.3 & 3745.0 & 20312.9 & 8926.2 \\
NGC4594 & $1.0\times 10^{9}$ & 9.2 & 2678.6 & 14509.2 & 6375.9\\
NGC4486B(M104) & $5.7\times 10^{8}$ & 15.3 & 1526.8 & 8470.3 & 3634.2 \\
NGC4261 & $4.5\times 10^{8}$ & 27.4 & 1205.4 & 6529.2 & 2869.1
\\
NGC7052 & $3.3\times 10^{8}$ & 58.7 & 883.9 & 4788.1 & 2104.0 \\
NGC4342(IC3256) & $3.0\times 10^{8}$ & 15.3 & 803.6 & 4352.8 & 1912.8\\
NGC3377 & $1.8\times 10^{8}$ & 9.9 & 482.1 & 2611.7 & 1147.7 \\
NGC0221(M32) & $3.4\times 10^{6}$ & 0.7 & 9.1 & 49.3 & 21.7 \\
NGC0224(M31) & $3.0\times 10^{7}$ & 0.7 & 80.4 & 435.3 & 191.3\\\hline

\end{tabular}
\caption{\label{tab:{table2}} Time delay between first and second
relativistic images for the black hole at the center of different
galaxies in case of Schwarzschild spacetime or $f(R)$ spacetime.
The global monopole parameter is vanished here. All the masses and
distances are taken from [7, 22, 39].}
\end{center}

\end{table}

\section{Discussion}

In this paper, we analyzed the gravitational lensing in the strong
field limit for the Schwarzschild black hole spacetime with a
solid deficit angle owing to a global monopole in the context of
$f(R)$ gravity theory which produce one cosmological boundary as a
result from that the expansion of the Universe is currently
undergoing a period of acceleration.  We employ several kinds of
elliptic integrals to show the deflection angle and time delay and
further discuss the integrals in the limiting case to reveal the
dependence of the large angle and time difference on the spacetime
structure and the generalization of the standard general
relativity. We find the $f(R)$ correction has significant effects
on the gravitational lensing. For violating the asymptotic
flatness, the spherically symmetric spacetime will not allow any
particles to propagate from or to infinity. We command thus that
the distance between the source and the lens should be restricted
among the radius of the photon sphere and the radius of the
cosmological horizon, $r_{0}<D_{SL}<r_{c}$, and so does the
distance between the lens and the observer. If the minimum
approach is close enough to the radius of the photon sphere, the
photon will wind around the black hole for several times before
escaping.
 This phenomenon is well known as the gravitational lensing in strong field limits.

We present the analytic expressions of the deflection angle
$\alpha$ and the time delay between relativistic images
$\bigtriangleup T_{n,m}$ in strong field approximation, and
relationships between the geometry and the observables such as the
angular separation $s$, the magnification of relativistic images
${\mathcal{R}}_{m}$, the compacted images position
$\theta_{\infty}$ and the minimum impact parameter $y_{m}$. We find the deviation from the standard general relativity
enhance the effect of gravitational lensing. All of the deflection
angle, the angular separation between the first image and the
compacted images, the minimum impact parameter and the time delay
increase as the increasing of dimensionless variable $GM\psi_{0}$. We find the time delay between first two relativistic images for Schwarzschild spacetime dominated by $f(R)$ gravity can be serval times larger than the time lag for Schwarzschild lens. Considering $f(R)$ lens located in the center of IC3256 as an example, the time delay of first and second images is more than 3 days if the derivation of the standard general relativity is large enough to $GM\psi_{0}=0.12$.
The effect from the correction to the original gravity
theory is evident which provides us a way to confirm whether the
Einstein's general relativity needs to be generalized.

\vspace{1cm}
\noindent \textbf{Acknowledge}

This work is supported by NSFC No. 10875043.

\newpage
\begin{figure}
\setlength{\belowcaptionskip}{10pt} \centering
  \includegraphics[width=15cm]{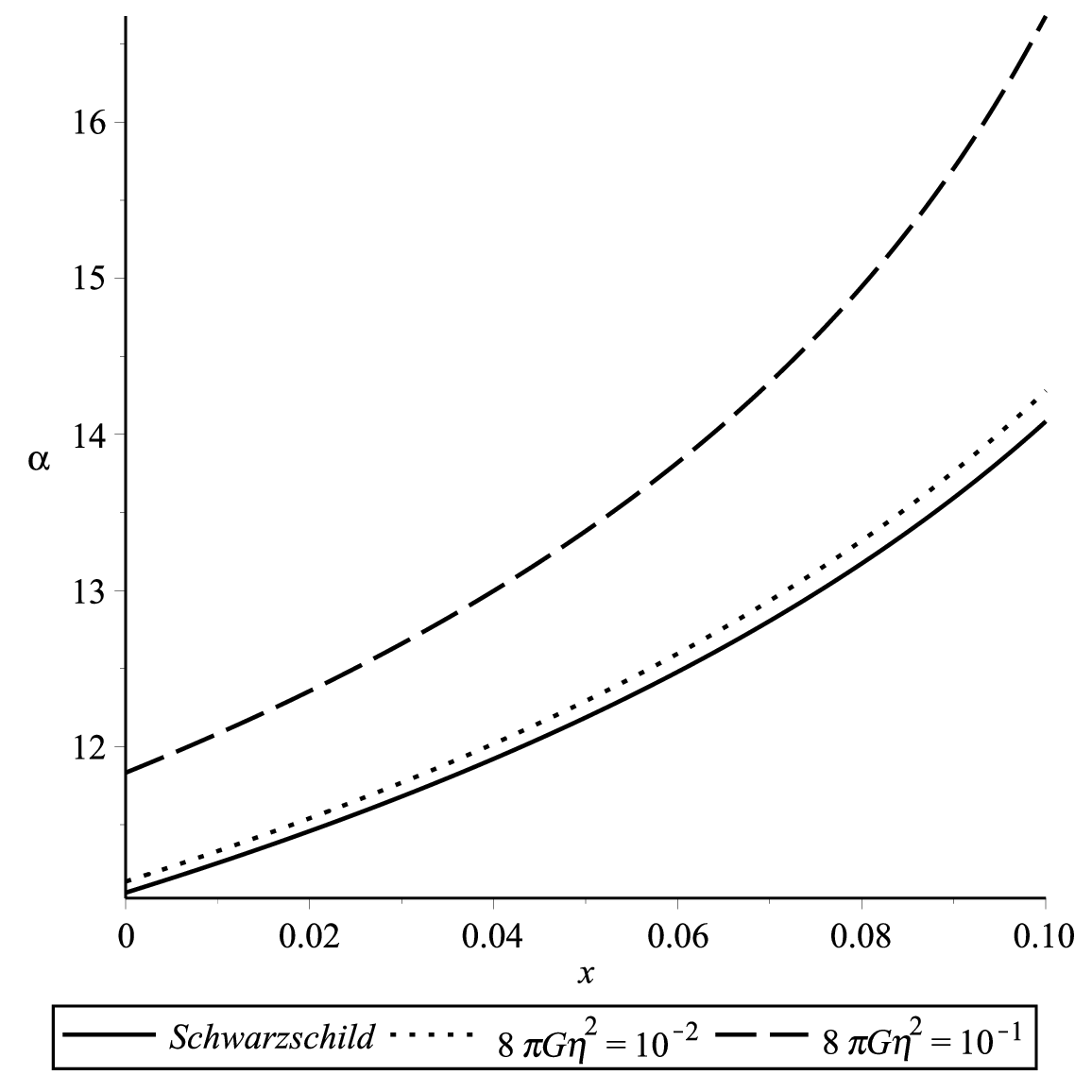}
  \caption{The dependence of deflection angle on $f(R)$
  parameter with variation $8\pi G\eta^{2}=10^{-5},10^{-2},10^{-1}$
  in the strong field limit with $y=y_{m}+0.003GM$.}
\end{figure}

\newpage
\begin{figure}
\setlength{\belowcaptionskip}{10pt} \centering
  \includegraphics[width=15cm]{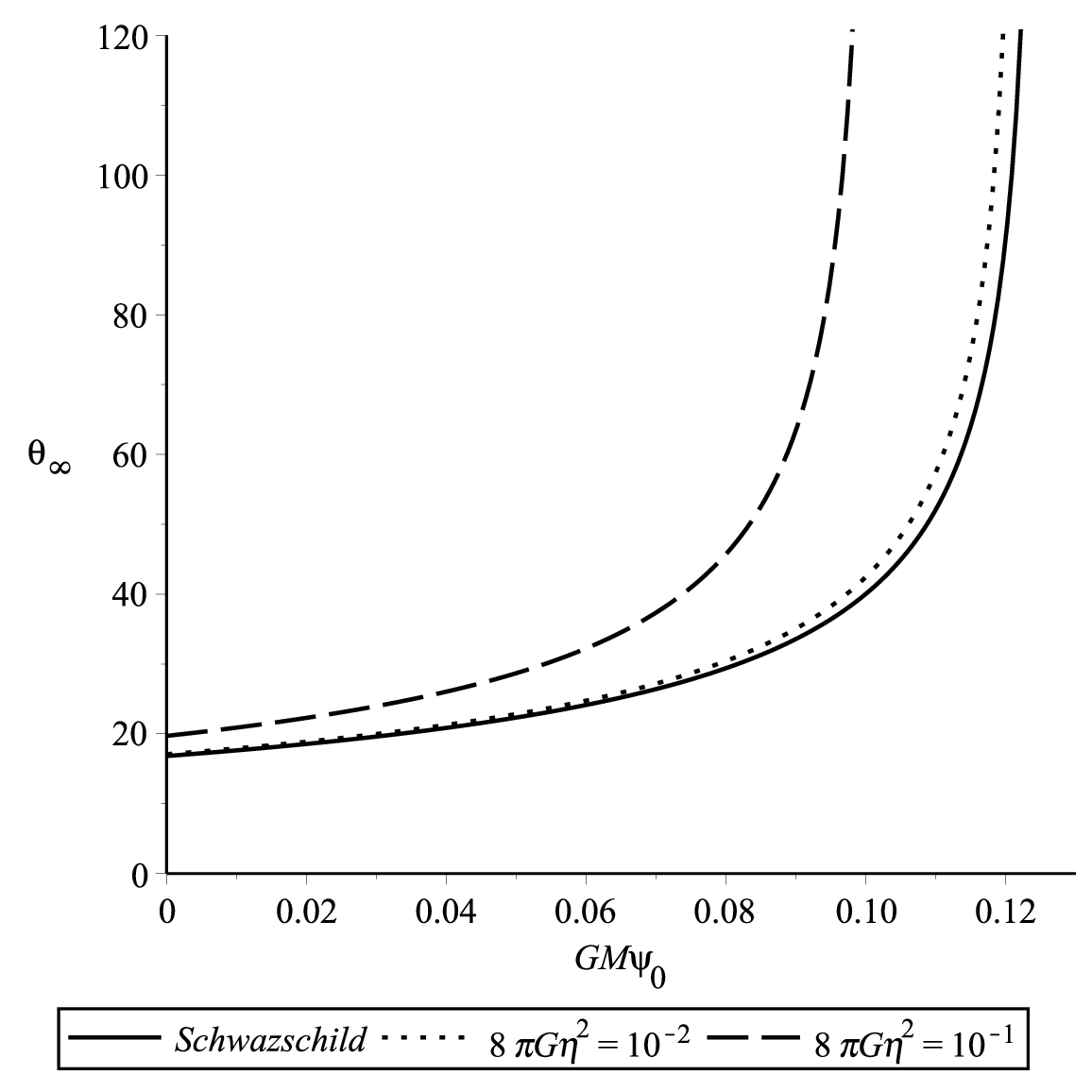}
  \caption{The behavior of the compacted images position $\theta_{\infty}$ ($\mu$ $ arc$ $sec$)
  on the dependence of the $f(R)$ parameter as $8\pi G\eta^{2}=10^{-5},10^{-2},10^{-1}$
  for varying the dimensionless modification parameter $GM\psi_{0}$.
  And we assume the black hole located in our galactic center,
  so $M=2.8\times 10^{6} M_{\bigodot}$ and $D_{OL}=8.5kpc$.}
\end{figure}

\newpage
\begin{figure}
\setlength{\belowcaptionskip}{10pt} \centering
  \includegraphics[width=15cm]{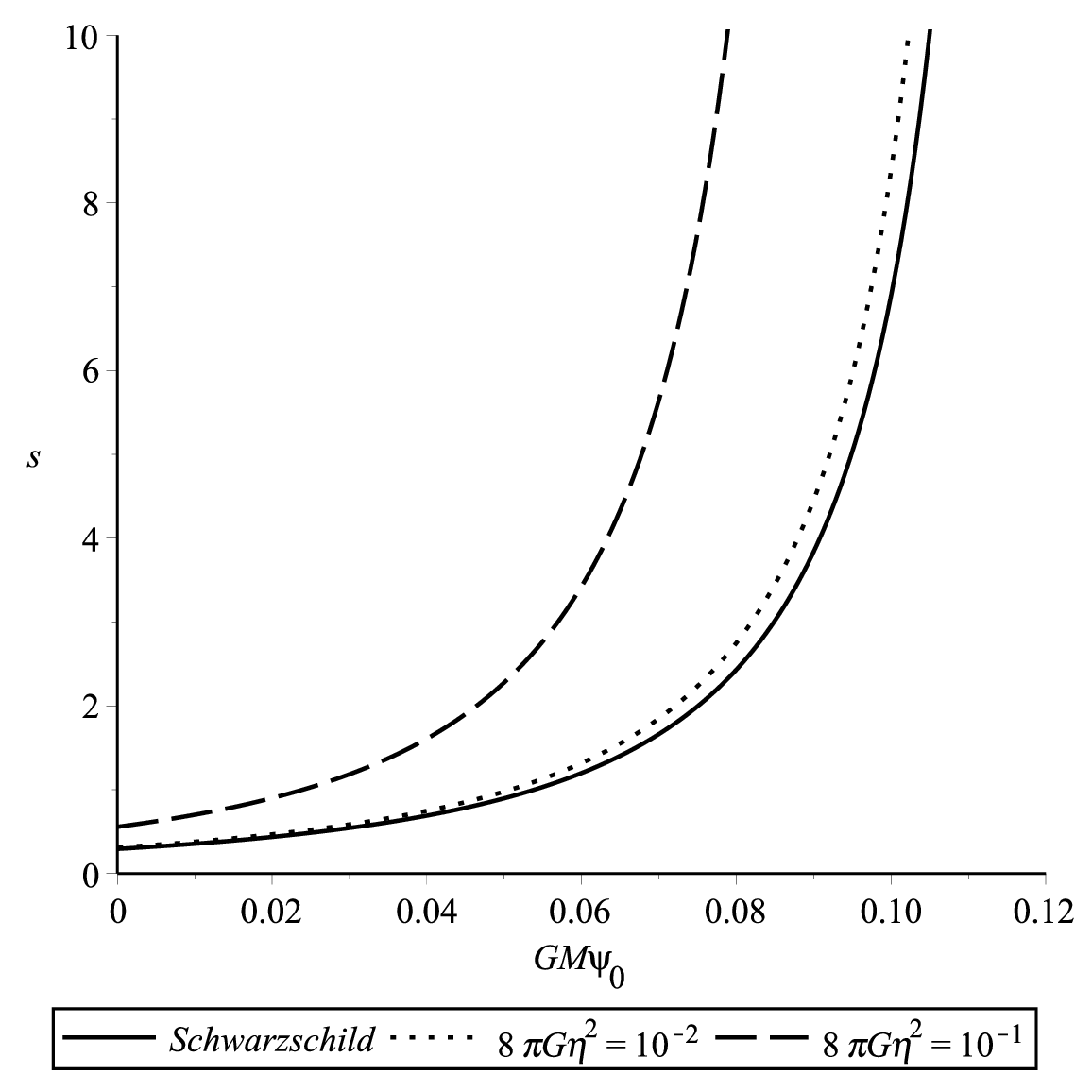}
  \caption{The figure shows the angular separation between the first image and the other
  compacted images $s$ ($\mu$ $ arc$ $sec$) as increasing functions of
  $GM\psi_{0}$. Here $M=2.8\times 10^{6} M_{\bigodot}$ and $D_{OL}=8.5kpc$.}
\end{figure}

\newpage
\begin{figure}
\setlength{\belowcaptionskip}{10pt} \centering
  \includegraphics[width=15cm]{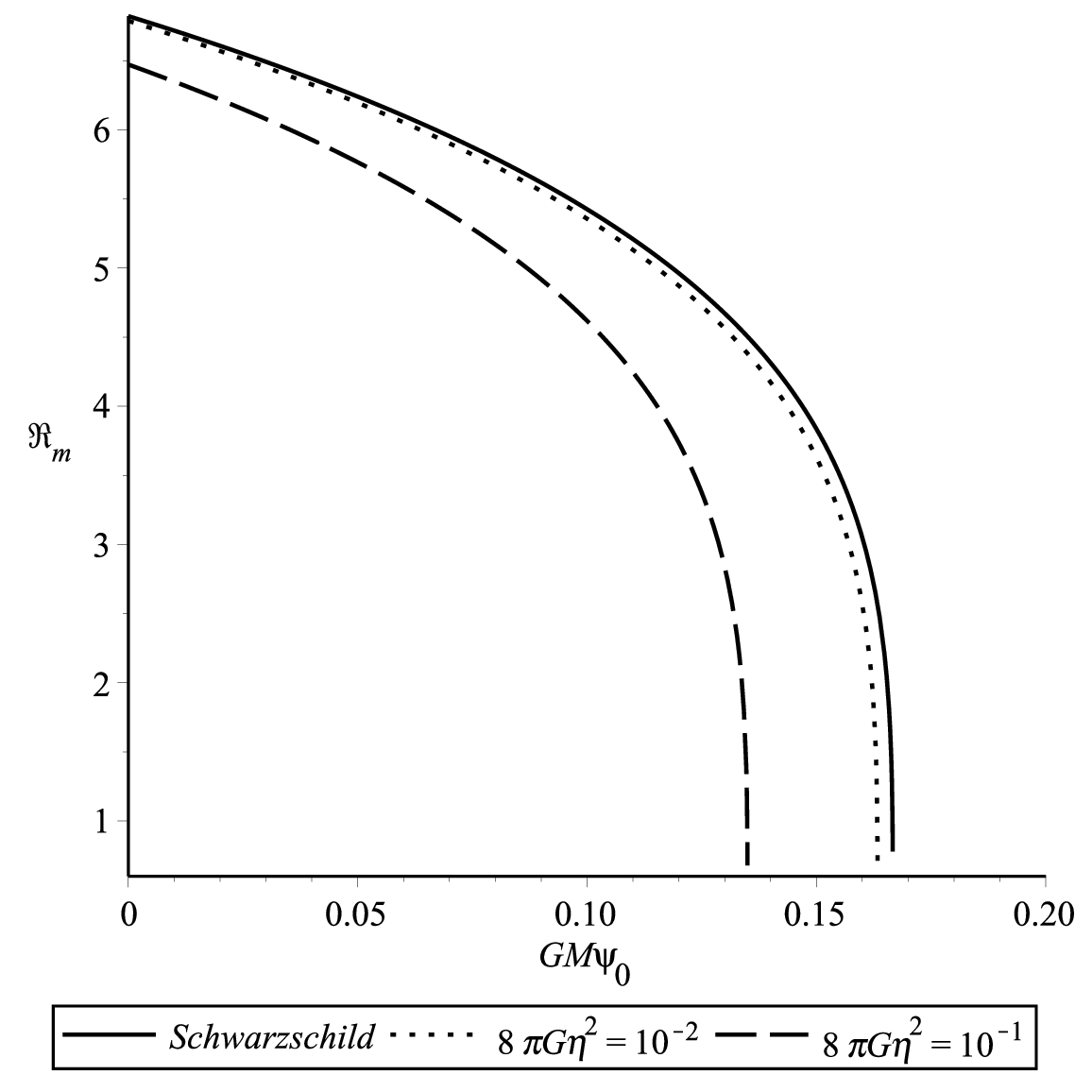}
  \caption{The figure shows the magnification of the first image and the all the other images
  ${\mathcal{R}}_ {m}$ as decreasing functions of $GM\psi_{0}$. The relationship between the ratio of the flux from the first image and the flux of all the other images and the magnification of the first image with the other images is ${\mathcal{R}}_{m}=2.5$log${\mathcal{R}}$. }
\end{figure}

\newpage
\begin{figure}
\setlength{\belowcaptionskip}{10pt} \centering
  \includegraphics[width=15cm]{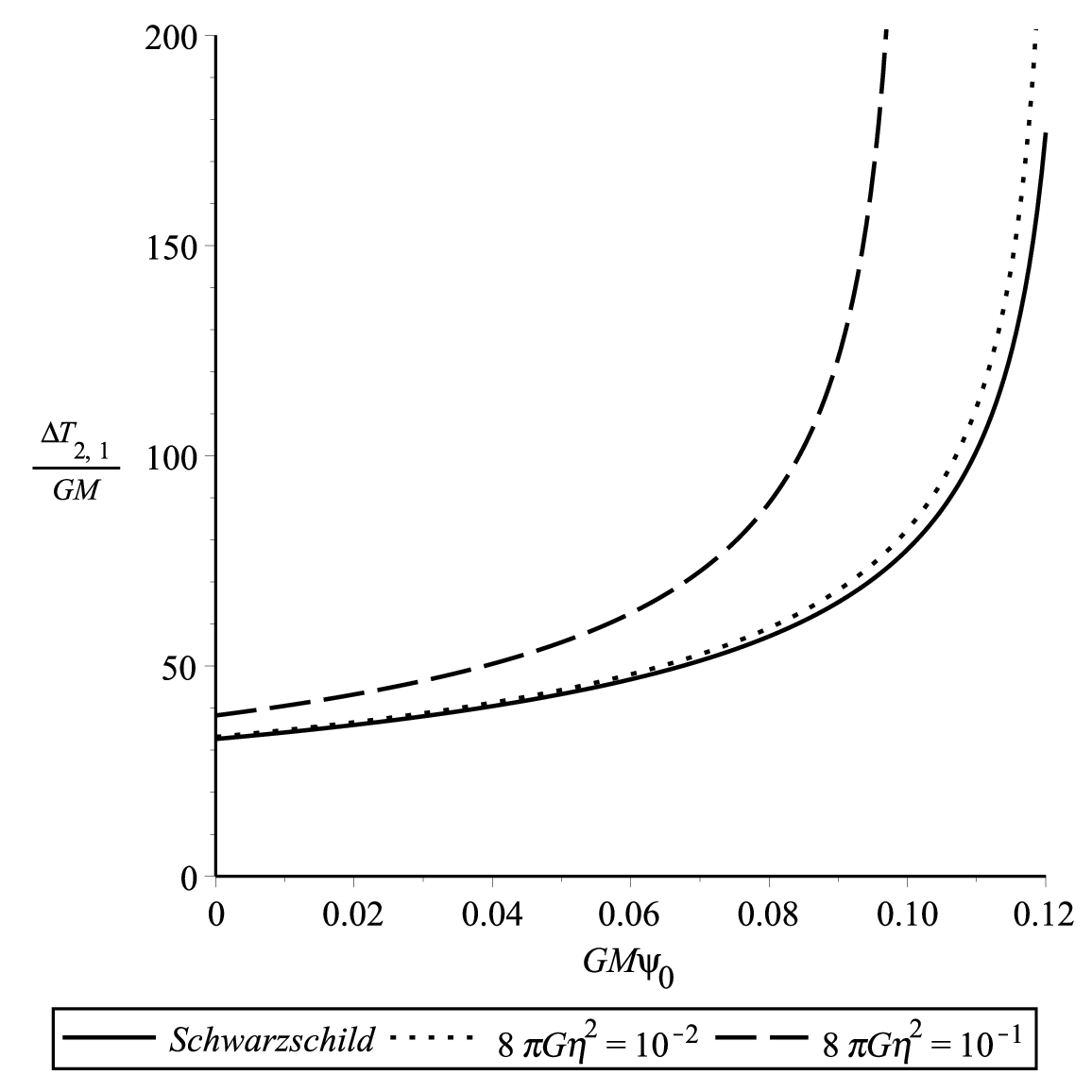}
  \caption{The figure shows the time delay between the first and
  the second relativistic images as increasing functions of $GM\psi_{0}$.}
\end{figure}

\end{document}